\documentclass[sigconf]{acmart}
\usepackage[normalem]{ulem}
\usepackage{units}
\usepackage{array}
\usepackage{multirow}
\usepackage{subfigure}
\AtBeginDocument{%
  }

\usepackage{comment}

\setcopyright{acmlicensed}
\copyrightyear{2025}
\acmYear{2025}
\acmDOI{XXXXXXX.XXXXXXX}

\acmConference[ICNS3 '25]{Make sure to enter the correct
  conference title from your rights confirmation emai}{August 19--21,
  2025}{Japan}

\acmISBN{978-1-4503-XXXX-X/18/06}

\begin{document}
\title{Coexistence analysis of Wi-Fi 6E and 
 5G \mbox{NR-U} in the 6 GHz band}

\author{Navid Kesthiarast}
\email{navid.keshtiarast@mcc.rwth-aachen.de}
\affiliation{%
  \institution{RWTH Aachen University}
  \city{Aachen}
  \country{Germany}}

  \author{Marina Petrova}
\email{petrova@mcc.rwth-aachen.de}
\affiliation{%
  \institution{RWTH Aachen University}
  \city{Aachen}
  \country{Germany}}

\begin{abstract}
The ever-increasing demand for broadband and IoT wireless connectivity has recently urged the regulators around the world to start opening the 6 GHz spectrum for unlicensed use. These bands will, for example, permit the use of additional \mbox{1.2 GHz}  in the US and 500 MHz in Europe for unlicensed radio access technologies (RATs) such as \mbox{Wi-Fi} and 5G New Radio Unlicensed (5G \mbox{NR-U}). To support QoS-sensitive applications with both technologies, fair and efficient coexistence approaches between the two RATs, as well as with incumbents already operating in the 6 GHz band, are crucial. In this paper, we study through extensive simulations the achievable mean downlink throughput of both \mbox{Wi-Fi} 6E APs and 5G \mbox{NR-U} gNBs when they are co-deployed in a dense residential scenario under high-interference conditions. We also explore how different parameter settings e.g., MAC frame aggregation, energy detection threshold and maximum channel occupancy time (MCOT) affect the coexistence.  Our findings give important insights into how to tune the key parameters to design fair coexistence policies. 
\end{abstract}


\maketitle
\section{Introduction}
\label{sec:Introduction}
\mbox{Wi-Fi} is undoubtedly the most prevalent wireless technology today, with more than \mbox{21.1 billion} devices in use \cite{WiFiGen}. It owes its success to the operation in unlicensed spectrum in the 2.4 GHz and \mbox{5 GHz} bands, which allowed for easy and quick entry of new players to develop innovative use cases. However, with the increased demand for high-throughput, high-reliability, and low-latency \mbox{Wi-Fi} connectivity due to applications such as wireless virtual and augmented reality (VR/AR), gaming, and holographic video, the need for new unlicensed spectrum and its importance have become more pronounced.

 To sustain the growing demand for wireless services, regulators across the Atlantic have taken decisive steps in the 6 GHz band. In the US, the Federal Communications Commission (FCC) has maintained its decision to open up the entire 5.925-7.125 GHz band for unlicensed use, a move that supports extensive Wi‑Fi innovation and unlicensed broad-spectrum access\cite{FCCApril2020, ITU2023}. In Europe, however, the regulatory approach is more segmented. The European Commission has reached an EU‑wide decision to make the lower portion of the band (5.945–6.425 GHz) available for wireless access—primarily to support enhanced \mbox{Wi-Fi} services, while planning to allocate the upper part of the band for commercial mobile use as part of a phased strategy\cite{ITU2023}. In the UK, Ofcom’s latest consultation outlines a proposal to enable standard‑power \mbox{Wi-Fi} in the lower 6 GHz under the control of an Automated Frequency Coordination (AFC) database, with a subsequent phase that introduces commercial mobile access in the Upper 6 GHz (6.425–7.125 GHz) via a prioritized spectrum split once European harmonization is finalized\cite{ofcom2025}. The IEEE 802.11 Working Group has already defined operating rules for \mbox{Wi-Fi} in this band through the 802.11ax (\mbox{Wi-Fi 6E}) standard\cite{Committee2007} and is advancing further with 802.11be, while 3GPP Release 16 and later define the operation of a 5G New Radio (NR) in the unlicensed band, called \mbox{NR-U}, which also foresees deployment of \mbox{NR-U} in a standalone mode\cite{release16}. This opens the opportunity of \mbox{NR-U} being deployed into scenarios in which \mbox{Wi-Fi} is already widely in use\cite{Sathya}, e.g., residential or campus scenarios. However, the disparity in the medium access control (MAC) parameters between these two radio technologies, such as detection thresholds and channel occupancy time, will inevitably result in unfair channel usage and significant network degradation. Therefore, ensuring fair spectrum sharing and harmonious coexistence between them is crucial.
\newline

\indent The issue of coexistence between \mbox{Wi-Fi} and 3GPP cellular technologies in the unlicensed bands came up for the first time and was extensively debated and studied in the context of LTE license-assisted access (LTE-LAA) in the 5 GHz band\cite{Andra_JSAC}. While the LTE-LAA had the constraint of being a new entrant in the 5 GHz band,  where already large number of \mbox{Wi-Fi} devices were operating, in the 6 GHz band both \mbox{NR-U} and \mbox{Wi-Fi 6E} are newcomers. This opens a window of opportunities to explore new coexistence approaches \cite{Naik2021}.
\newline
\indent While the coexistence between LTE-LAA and \mbox{Wi-Fi} is well understood \cite{Voicu2019}, studies on the coexistence of \mbox{NR-U} and \mbox{Wi-Fi} have recently started to emerge..
The survey articles by Sathya \textit{et al.} \cite{Sathya} and Naik\textit{ et al.} \cite{Naik} provide a comprehensive overview of the proposed rules, challenges and standardization efforts for enabling coexistence in 6 GHz. In \cite{Naik2021}, the authors use stochastic geometry to study the uplink performance of 802.11ax when coexisting with \mbox{NR-U} in the \mbox{6 GHz} bands. 
In contrast, we analyse the downlink performance of both technologies with link-level simulations to capture the interference and the interactions more realistically. The work in \cite{Luo2022} analyses coexistence using 3GPP fairness criteria for the 5 GHz band, where \mbox{Wi-Fi} is already deployed and regarded as an incumbent technology. However, in the \mbox{6 GHz} band, neither \mbox{Wi-Fi} nor \mbox{NR-U} is incumbent technology, necessitating different criteria for harmonisation coexistence. Therefore, in our work, we present an end-to-end coexistence simulation framework for the 6 GHz band in residential scenarios. In our analysis, we address the following questions: \mbox{(i)} what is the effect of different MAC parameters on the overall mean throughput of \mbox{Wi-Fi} and \mbox{NR-U}, and \mbox{(ii)} how is the fairness criterion maintained in the coexistence scenario without biasing towards one technology?

The contributions of this paper are as follows:
\begin{itemize}
 \item  We investigate the coexistence of \mbox{Wi-Fi} 6E and 5G \mbox{NR-U} in a residential scenario, focusing on mean achievable downlink throughput and fairness. We use ns-3 simulator to analyze the coexistence between \mbox{NR-U} and \mbox{Wi-Fi} 6E in the 6 GHz band. Our extensive link-level simulations provide valuable insights into the performance of both technologies and their mutual influence.
\item We present an analytical framework to calculate the mean downlink throughput in the coexistence scenario.
We then validate the mean downlink throughput results obtained with the ns-3 NR-U framework\cite{Patriciello} for sub-7 GHz against our analytical framework.
\item Finally, we assess the impact of key parameters such as frame aggregation, energy detection thresholds, and maximum channel occupancy time (MCOT) on the throughput performance and fairness when both technologies are competing for the same channel. Our study highlights the importance of tuning these parameters to the right values to achieve fair coexistence between \mbox{NR-U} and \mbox{Wi-Fi} 6E.
\end{itemize} 


The remainder of the paper is organized as follows. Section \ref{Background} provides a brief background of Wi-Fi 6E and 5G \mbox{NR-U}. In Section \ref{System_Model_and_Methodology}, we discuss the system model, parameters and simulation scenario. In Section \ref{Throughput_Model_sec}, we present the throughput model analysis. Section \ref{sec:SIMULATION RESULTS AND DISCUSSION} presents and discusses our simulation results. In Section \ref{Conclusions}, we conclude the paper.
\section{Background}
\label{Background}
\subsubsection*{\textbf{Wi-Fi 6E}}
Wi-Fi 6E is the extension of Wi-Fi 6 (IEEE 802.11ax) in the recently unlocked 6 GHz bands. Wi-Fi 6E devices take advantage of the large amount of newly available spectrum and enjoy higher capacity, higher speeds, and lower latency. The 59 newly available 20 MHz channels in the US (24 in Europe) will drastically reduce the congestion. This is particularly beneficial for high-density networks e.g.,  convention centers, shopping malls, and stadiums. Moreover, the increased number of wider non-overlapping channels of 40, 80 and 160 MHz will allow for multi-gigabit Wi-Fi speeds for end users.      

Just like Wi-Fi 6, Wi-Fi 6E uses Orthogonal Frequency Division Multiple Access (OFDMA). OFDMA divides the channel into sub-channels, also known as Resource Units (RU), and allows multiple users to communicate simultaneously, rather than waiting for their turn, which is the case in the older standards (e.g., IEEE 802.11 n/ac). Both downlink and uplink multi-user OFDMA (MU OFDMA)
transmissions are initiated by the Wi-Fi 6 Access Point
(AP) using a so called Trigger Frame (TF). It is important to note that Wi-Fi 6E implements OFDMA on top of CSMA-CA, meaning that, an AP has to first contend for the channel with other coexisting technologies before transmitting the TF. Using TF, the AP schedules the uplink transmissions of all addressed stations (STAs) and sends the required resource allocation information. Before starting the transmission on the allocated RUs in the uplink, the STAs have to sense the channel idle. If the channel is busy, the STAs have to refrain from transmission. If the channel is idle, the STAs transmit immediately without performing back-off.

\subsubsection*{\textbf{5G \mbox{NR-U}}}
5G New Radio-Unlicensed is a RAT that is being developed by 3GPP and first introduced in Release 16\cite{release16}. In the design of \mbox{NR-U}, the LTE-LAA, an LTE version standardized in Release 13 \cite{RP182878} and designed to operate in the 5 GHz unlicensed bands along with Wi-Fi, has been used as a baseline. One of the key inherited features is the channel access mechanism, namely Listen Before Talk (LBT). This means that before transmission, \mbox{NR-U} devices have to sense the channel to ensure harmonious coexistence with other unlicensed devices, such as IEEE 802.11ax.

Building on these foundations, Release 18\cite{3gpp2024} introduces further refinements to NR‑U’s channel access procedures by defining two distinct categories: Type 1 and Type 2 Channel Access Procedures. We use type 1 category in this work and will refer to it as the LBT Procedure. The procedure operates as follows: after a quiet period of $16~\mu s$, the gNB performs a channel check by undergoing $d_i$ successive sensing slots, each lasting $9~\mu s$. The overall deferment time, $T_{df}$, is determined by adding $T_f$ to the product of $d_i$ and $T_{CCA}$, with the value of $d_i$ chosen according to the priority assigned to each traffic type in the standard. If the channel remains free during this period, the gNB then enters a backoff phase by selecting a random number from the range $\{0,1,..., CW\}$ and decrementing it while monitoring the channel. When the counter reaches zero, transmission begins. Should the channel become active during any sensing slot, the backoff countdown pauses and resumes with the remaining count once the channel is clear. Upon gaining access, the gNB is allowed to transmit for a period limited by the Maximum Channel Occupancy Time (MCOT), which varies with different traffic types. The contention window ($CW$) is updated based on the feedback received from Hybrid automatic repeat request (HARQ). If at least 10\% HARQ-ACKs (for code block group feedback) are received, $CW$ resets to its minimum; otherwise, it increases to the next permitted value by doubling the $CW$.


5G \mbox{NR-U} adopts the 5G NR PHY layer and thus can benefit from its enhancements, such as flexible numerologies and mini-slot scheduling. Using mini-slots for scheduling the \mbox{NR-U} transmissions could solve the issue of needing a long reservation signal that LTE-LAA devices need to send, so that the transmission could start at the beginning of a sub-frame. We will treat the essence of reservation signal for coexistence in more detail in Section \ref{subsection:Simulation Tool}

In contrast to LTE-LAA, where the 5 GHz license-free band can only be used as a secondary channel to a licensed band, the \mbox{NR-U} can be operated independently on the license-free band. This opens the possibility of deploying local 5G cellular networks in the 6 GHz bands.
\section{System Model and Methodology}
\label{System_Model_and_Methodology}
\subsection{Setup and Scenarios}
\label{sec:ScenariosandSetup}
\begin{figure}[h!]
	\centering
	\includegraphics[width=0.5\textwidth,trim = 40mm 110mm 32mm 36mm,clip]{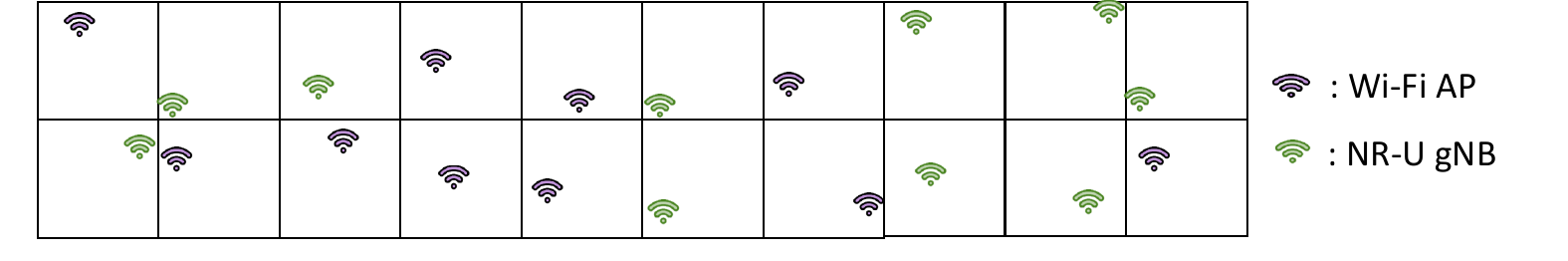}
	\caption{Residential scenario, with a random distribution of 10 AP and 10 gNBs.}
	\label{fig:layout2}
\end{figure}

We consider an indoor residential scenario, where  \mbox{Wi-Fi} 6E and 5G \mbox{NR-U} are independently deployed. The 5G gNBs and \mbox{Wi-Fi} APs\footnote{In this paper, we will use the terms AP, and \mbox{Wi-Fi} to indicate \mbox{Wi-Fi} 6E. Similarly, gNB and \mbox{NR-U} refer to 5G \mbox{NR-U}.} 
are distributed inside a single-floor building with 20 apartments. Each apartment is of size $10 m\times 10 m\times 3 m$. In the defined scenarios, \mbox{Wi-Fi} has 10 APs, whereas the number of gNBs varies from $0$ to $30$. The placement of the APs and gNBs is done as follows. 
Initially, we place 10 APs random, maximum one AP per apartment. In the other 10 apartments, gNBs are placed randomly. When all apartments have one AP/gNB, the rest of the gNBs are again distributed randomly. The number of AP/gNBs in each apartment is limited to two. Figure \ref{fig:layout2} shows an example scenario for 10 AP and 10 gNBs. 

\mbox{Wi-Fi} APs, like gNB, schedule uplink and downlink and assign resource blocks to user equipments (UEs) \cite{Naik}. Our study does not address UE resource block scheduling since the fairness between \mbox{NR-U} and \mbox{Wi-Fi} depends only on the channel access mechanism and contention method. The scheduling of resource blocks between two or more UEs is not the subject of this study.
For calculating the path loss, we use the multi-wall-and-floor (MWF) model, where the first wall causes an attenuation of 16 dB and the following traversed walls have an attenuation of 14 dB\cite{Lott2001}.
Since we consider one-floor building, we can write the MWF model for a transmitter and receiver at the distance of $d$ as follows:
\begin{equation} \label{eqnchannel}
 L_{MWF}(d)=L_0+10n \log{d}+\sum^{NoW}_{i=1}{\sum^{NTW}_{k=1}{L_{w_{ik}}}}
 \end{equation}
 where $L_0$ is the path loss at a distance of 1 m. $NoW$ and $NTW$ are the number of wall types and the number of traversed walls, respectively. $w_{ik}$ represents the attenuation of type $i$ and $k$-th traversed wall. Table~\ref{tab:parameters} lists all system parameters.
 
\begin{table}[t]
	\caption{System and Simulation Parameters}
	\begin{tabular}{|p{2.3cm}|p{5.6cm}|}
		\hline
		Network Size& 10 Wi-Fi APs, 0-30 gNBs
		\\
		\hline
		Carrier frequency& 5.955 GHz\\
		\hline
		Transmit power& 23 dBm
        \\
		\hline
		Channel bandwidth &20 MHz
		\\
		\hline
	    Frame Aggregation&  
	    No frame aggregation/ A-MSDU/ A-MPDU
		\\
		\hline
		Simulation duration& {4 s data transmission} 
		\\
		\hline
		Traffic& Downlink saturated best-effort traffic with a payload of 1474 bytes per packet
		\\
		\hline
	
		Sensing threshold&{ -62 dBm, -72 dBm, -82 dBm} 
		\\
		\hline
			MCOT&{ 8 ms, 5 ms} 
		\\
		\hline
			No. of simulations&{100 iterations for each network size}
		\\
		\hline
	\end{tabular}
	\label{tab:parameters}
\end{table}
\subsection{Simulation Tool}
\label{subsection:Simulation Tool}
We use ns-3 (v3.35) to conduct the simulations. The \mbox{NR-U} module (v1.2)\cite{Patriciello} is used to simulate \mbox{NR-U} gNBs. For simulating \mbox{Wi-Fi}, we use the ns-3 \mbox{Wi-Fi} module, which provides a packet-level abstraction of the PHY and an accurate MAC-level implementation of IEEE 802.11 ax standard. In addition, a rate control is enabled, using the Minstrel algorithm\cite{Minstrel} implementation in ns-3. This method adapts the rate based on probing the environment and calculating the probability of successful transmission. This allows the modulation and coding scheme (MCS) to adapt depending on the channel conditions.

As discussed in Section \ref{Background}, in the context of \mbox{NR-U}, the implementation of the mini-slot scheduling can reduce the necessity for a reservation signal. However a reservation signal remains essential to secure the channel in case of coexistence with Wi-Fi or another gNB. Without a reservation signal, Wi-Fi APs can occupy the channel even if the gNB has successfully completed the LBT procedure. This occurs because Wi-Fi APs can start the transmission immediately after finishing the CSMA/CA procedure, while NR‑U gNBs transmissions are synchronized to slot boundaries AND NR-U gNBs have to wait until the beginning of the next slot, which can result in a gap between the end of the backoff procedure and the start of the transmission slot. This is illustrated in Figure~2(b).
Alternatively, it may happen that two \mbox{NR-U} gNBs finish their backoff procedure in the previous slot, as shown in Figure~2(b). In the absence of a reservation signal, they will start transmission at the next slot simultaneously, which causes interference between the \mbox{NR-U} gNBs.
In \cite{NR_U_Type1}, the authors propose two solutions to bridge this gap: (i) incorporating additional sensing immediately before transmission, and (ii) delaying the start of the backoff process so that it concludes closer to the slot boundary. However, these approaches may lead to fairness issues since Wi‑Fi, which employs load-based scheduling, can initiate transmission immediately once the channel is detected as idle as illustrated in Figure~2(a). To address this challenge, our work employs a reservation signal after a successful backoff and channel acquisition, ensuring that the channel remains occupied until the slot boundary is reached as shown in Figure~2(c).  For reproducibility and to
further advance research in the field, we have made the simulation code publicly available at\cite{keshtiarast_NRU}.




\begin{figure}[!ht]
\centering
\begin{subfigure}
        \centering
        \includegraphics[width=0.56\textwidth,trim = 25mm 67mm 40mm 60mm,clip]{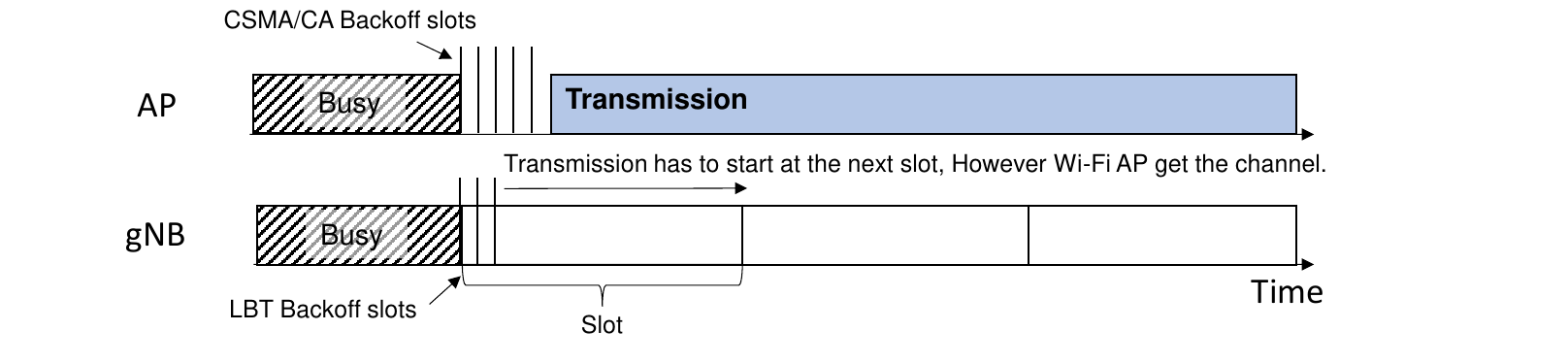}
        \vspace{-0.75cm}
        \caption*{(a)}
        \label{subfigreserv1}
        \end{subfigure}
\begin{subfigure}
        \centering
        \includegraphics[width=0.56\textwidth,trim = 25mm 60mm 40mm 60mm,clip]{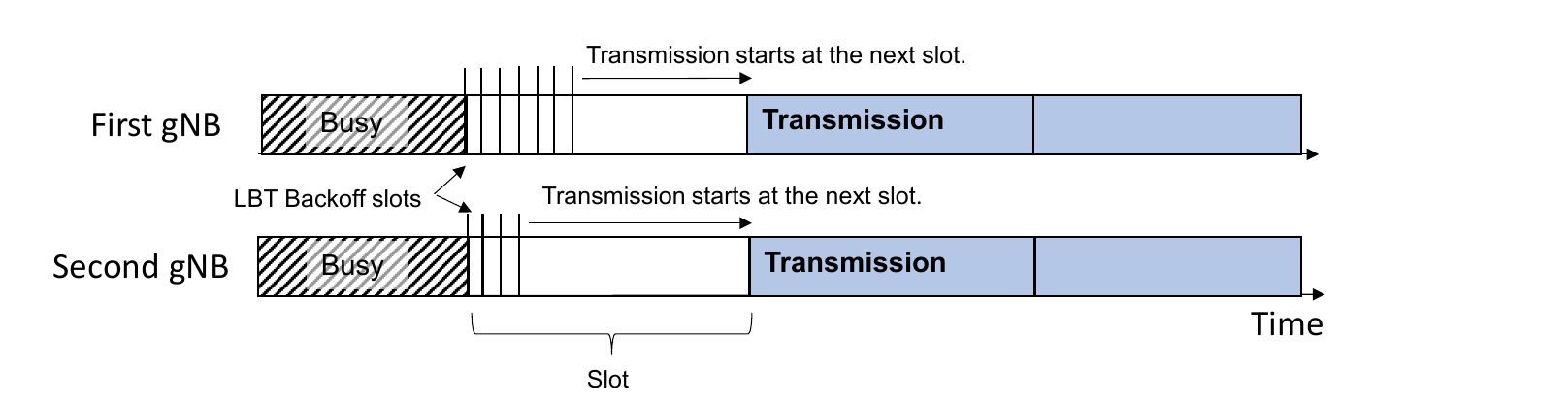}
        \vspace{-0.75cm}
        \caption*{(b)}
        \label{subfig:reserv2}
        \end{subfigure}
\begin{subfigure}
        \centering
        \includegraphics[width=0.56\textwidth,trim = 25mm 85mm 40mm 70mm,clip]{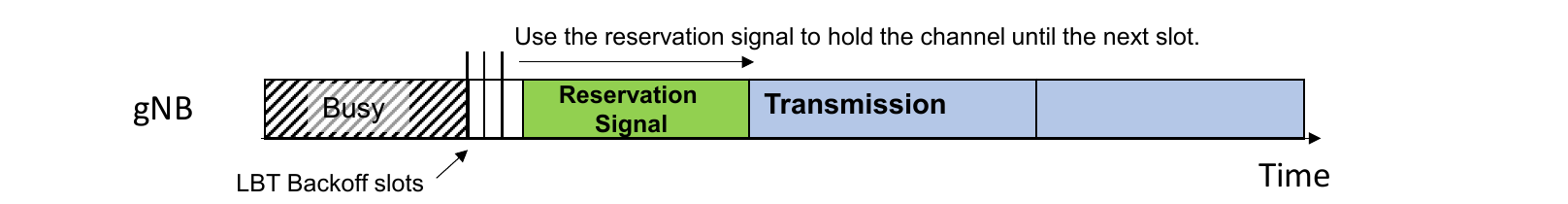}
        \vspace{-0.75cm}
        \caption*{(c)}
        \label{subfig:reserv3}
        \end{subfigure}        
        \caption{Lack of reservation signal causes interference: (a) between Wi-Fi APs and gNB, (b) between gNBs; (c) a reservation signal secures the channel until the gNB’s slot-aligned transmission begins.}
        \label{fig:NRSLOT}
\end{figure}

\subsection{Parameters influencing the coexistence}
 In the following, we briefly review several parameters of \mbox{Wi-Fi} and \mbox{NR-U} that affect the coexistence in terms of throughput and fair channel access. The results and the analyses of their influence on the coexistence are shown later in Section~\ref{sec:SIMULATION RESULTS AND DISCUSSION}.
\subsubsection{Frame Aggregation Types}
Frame aggregation is a method of combining multiple frames into a single frame transmission. It is one of the key enablers of more efficient airtime usage, and it has a direct influence on the \mbox{Wi-Fi} transmission duration time. The IEEE 802.11ax standard defines two types of frame aggregation, namely:
\begin{itemize}
\item Aggregate MAC Service Data Unit (A-MSDU): By using A-MSDU, several payload units are aggregated together with one MAC header.        
\item Aggregate MAC Protocol Data Unit (A-MPDU):
In A-MPDU, the MAC header of each frame is preserved, and the whole frames are aggregated together.
\end{itemize}
In this work, all \mbox{Wi-Fi} APs use the same frame aggregation type in each simulation.
The \mbox{Wi-Fi} APs apply one of the following configurations for frame aggregation (a) no frame aggregation, (b) A-MSDU with fixed maximum available size in standard (11398 octets), and (c) A-MPDU with fixed maximum available size in the standard (6500631 octets).
Based on IEEE 802.11ax standard \cite{Committee2007}, the maximum transmission time is limited to 5.484 ms. Using frame aggregation enables \mbox{Wi-Fi} APs to retain the channel for a longer period of time without repeating the CSMA operation, consequently enhancing \mbox{Wi-Fi} throughput performance. 
\subsubsection{Different Sensing Thresholds}
\mbox{Wi-Fi} devices in the \mbox{5 GHz} bands use two carrier sensing thresholds to identify and defer to other \mbox{Wi-Fi} devices and non-\mbox{Wi-Fi} technologies like LTE-LAA.
IEEE 801.11ax sensitivity varies on MCS and channel bandwidth. The lowest carrier detection threshold for 20 MHz bandwidth is -82 dBm, however, non-\mbox{Wi-Fi} frames are detected at -62 dBm. Nonetheless, \mbox{NR-U} devices employ -62 dBm as the energy detection (ED) threshold for detecting all technologies. Given that inherent asymmetry in the ED thresholds used by LTE-LAA and \mbox{Wi-Fi} is one of the root causes of poor coexistence between LTE and \mbox{Wi-Fi} in 5 GHz \cite{Sathya}, we investigate the impact of both common and asymmetric ED thresholds for both \mbox{NR-U} and \mbox{Wi-Fi} devices on throughput performance. In our analysis, we consider ED threshold values of -62 dBm, -72 dBm, and -82 dBm. Specifically, we set the same ED threshold for all APs and gNBs and examined for the three ED threshold values described above. Then, we set the ED of the APs to -62dBm and varied the ED values of the gNBs to -72 dBm and -82 dBm. 
\subsubsection{Maximum Channel Occupancy Time (MCOT)}
MCOT can affect channel access fairness. After LBT and channel access, \mbox{NR-U} gNB can use the channel for up to MCOT for downlink and uplink traffic. In the coexistence scenario, MCOT adjustments can affect \mbox{NR-U} airtime and throughput fairness. In this paper, we investigated the various sensing thresholds using an MCOT of 8 ms from 3GPP specification\cite{release16} for best-effort traffic.
As the maximum transmission time for \mbox{Wi-Fi} AP is 5.484 ms, we also investigate the impact of using an MCOT of 5 ms for the gNBs in Section \ref{MCOT_impact}. 
\section{Throughput Model}
\label{Throughput_Model_sec}
In this section, we analyze the throughput that each AP/gNB can achieve in a coexistence scenario, leveraging the model introduced in \cite{Andra_JSAC}. We enhance the throughput model's precision by incorporating the reservation signal.
As both \mbox{NR-U} and \mbox{Wi-Fi} technologies employ LBT as medium access, the throughput of AP/gNB $x$ is 
\begin{equation} \label{eqn1}
	R_x = {S_x \times AirTime_x {\times} {\rho}_x (SINR_u)} 
\end{equation}
where $S_x$ denotes the MAC efficiency, $AirTime_x$ represents the fraction of total airtime used for successful transmissions and collisions by AP/gNB $x$. Further, ${\rho}_x(\cdot)$ represents the mapping function and $SINR_u$ denotes the received SINR at the user. The expressions for all the parameters are explained in detail below.

The MAC efficiency of AP/gNB $x$, $S_x$, is defined as the fraction of successful packet transmission time. Both \mbox{Wi-Fi} and \mbox{NR-U} use carrier sense before the transmission of packets. The overall time is divided into slots, and each slot can either be (i) an idle slot of duration $\sigma$, (ii) an average successful transmission duration of $\overline{\rm{T_{s,x}}}$, (iii) average collision duration of $\overline{\rm{T_{c,x}}}$. Therefore, following the analysis of Bianchi’s model \cite{Bianchi}, the expression for $S_x$ is,
\begin{equation} \label{eqn2}
S_x=\frac{\overline{\rm {T_{f,x}}}}{\overline{\rm {T_{s,x}}}-\overline{\rm {T_{c,x}}}+ \sigma \frac{\overline{\rm {T_{c,x}/\sigma}}-(1-\tau)^n(\overline{\rm{T_{c,x}/\sigma}}-1)}{n\tau(1-\tau)^{n-1}}}
\end{equation}
where $n$ is the total number of APs and gNBs, and $\tau$ refers to the likelihood that a station will transmit during a randomly selected time slot. Further, $\overline{\rm {T_{f,x}}}$ is the average frame duration in the sensing range, which is calculated by taking the average frame duration of AP/gNB x and all nodes in the sensing range, which means
\begin{equation}\label{eq_20}
\overline{T_{f,x}}=\frac{T_{f,x}+ \displaystyle \sum_{z\in\mathbf{A}_x}T_{f,z} + \sum_{z\in\mathbf{B}_x}T_{f,z}}{1+|\mathbf{A}_x|+|\mathbf{B}_x|}
\end{equation}
\noindent In our model, populations A and B represent the \mbox{Wi-Fi} and \mbox{NR-U} technology. $|\mathbf{A}_x|$ and $|\mathbf{B}_x|$ are the numbers of APs and gNBs in the sensing range, respectively. Further, the duration of the frame depends on the technology given below:
\begin{equation}\label{eq_23}
T_{f,x}=\begin{cases}
   \begin{split} PHY&_{header}+MAC_x, \text{if $x$ is Wi-Fi 6E AP} \end{split}\\
   \begin{split} MCOT-\mathbb{E}[T_r]&, \text{if $x$ is NR-U gNB} \end{split} 
   \end{cases}
\end{equation}
where $MAC_x = \frac{MAC_{header}+MSDU}{R_x}$. \mbox{Wi-Fi} frame duration depends on the MAC header, MAC service data unit size and the data rate ($R_x$). While for \mbox{NR-U}, it depends on the time difference between MCOT and the reservation signal duration ($T_r$). We assume that the length of the reservation signal follows a uniform distribution over the interval [0 ~$\Delta$] where $\delta$  is dependent on the duration of the mini-slot. This duration, determined by the number of included OFDM symbols, can take on possible values from the set $\{9,18,36,63,126,250,500,1000\}\mu s$.

The average successful transmission duration $\overline{T_{s,x}}$ of AP/gNB $s$ in the sensing range is given as follows:
\begin{equation}\label{eq_20_24}
\overline{T_{s,x}}=\frac{T_{s,x}+ \displaystyle \sum_{z\in\mathbf{A}_x}T_{s,z} + \sum_{z\in\mathbf{B}_x}T_{s,z}}{1+|\mathbf{A}_x|+|\mathbf{B}_x|}
\end{equation}
where the duration of successful transmission is equal to
\begin{equation}\label{eq_24}
T_{s,x}=\begin{cases}
    \begin{split} T_{f,x}&+DIFS+SIFS+PHY_{header}+\frac{ACK}{R_{x,min}}, \\&\text{if $x$ is Wi-Fi 6E AP}  \end{split}\\
    \begin{split} T_{f,x}&+T_d+\mathbb{E}[T_r], &\text{if $x$ is NR-U gNB}  \end{split}
    \end{cases}
\end{equation}
In case of \mbox{Wi-Fi}, $T_{s,x}$ includes frame transmission duration, interframe spacings (both DIFS and SIFS), physical layer headed and acknowledgement frame transmission duration. Further, in the case of \mbox{NR-U}, the successful transmission duration includes defer time, $T_d$, the reservation signal duration and frame transmission duration. The values of $SIFS$ and $DIFS$ are equal to $16\mu$s and $34\mu$s, respectively. $T_d$ is equal to $34\mu$s.

Similarly, the average collision duration in the sensing range and the collision duration are defined as follows:
\begin{equation}\label{eq_20}
\overline{T_{c,x}}=\frac{T_{c,x}+ \displaystyle \sum_{z\in\mathbf{A}_x}T_{c,z} + \sum_{z\in\mathbf{B}_x}T_{c,z}}{1+|\mathbf{A}_x|+|\mathbf{B}_x|}
\end{equation}
where the collision duration is equal
\begin{equation}\label{eq_25}
T_{c,x}=\begin{cases}
    \begin{split} T_{f,x}&+DIFS, \text{if $x$ is Wi-Fi 6E AP} \end{split}\\
    \begin{split} T_{f,x}&+T_d+\mathbb{E}[T_r], \text{if $x$ is NR-U gNB} \end{split}
    \end{cases}
\end{equation}
Further, $AirTime_x$ in the equation \ref{eqn1} denotes how much time AP/gNB x acquires based on its MAC and the MAC parameters of other APs/gNBs in the sensing range, as follows:
\begin{equation} \label{eqn22}
AirTime_{x}=\frac{T_{f,x}\times{p_x}}{ T_{f,x}\times{p_x}+\sum_{z\in A_x,B_x}{T_{f,z}p_z}}
\end{equation}
$p_{x}$ is defined as the transmission probability, which is given by $p_{x}=\frac{1}{ 1+|A_x|+|B_x|}$.
Finally, ${\rho_x (SINR_u)}$ is the mapping function between the SINR and spectral efficiency of the \mbox{Wi-Fi}/\mbox{NR-U} \cite{Committee2007,release16}. 
The received SINR at user $u$ associated to AP/gNB $x$ is $SINR_u=\frac{P^t_x \times l}{I_{APs}+I_{gNBs}+N_0}$, where $l$ is the path loss between AP/gNB and the associated user. $I_{APs}$ and $I_{gNBs}$ denote the aggregated interference from APs and gNBs located outside the carrier sense range of AP/gNB $x$, which also employ LBT procedure. Therefore, the aggregate interference caused by nodes outside CS-range depends on their air time, and it can be calculated using the following equations

\begin{equation} \label{eqn4}
 \small
 I_{APs}=\sum_{z\in APs \setminus x}{\frac{P\times L_{u,z}^{-1}}{1+|A_z|+|B_z|}}
 \end{equation}
 \begin{equation} \label{eqn5}
 \small
 I_{gNBs}=\sum_{z\in gNBs \setminus x}{\frac{P\times L_{u,z}^{-1}}{1+|A_z|+|B_z|}},
 \end{equation}
 where $P$ is the transmitted power of the interferer and $L_{u,z}$is the path loss between the interferer and the user $u$.

In Section \ref{sec:SIMULATION RESULTS AND DISCUSSION}, we present the results of the mean throughput of \mbox{NR-U} and \mbox{Wi-Fi} networks by using the throughput model and running Monte Carlo network realizations. We also present the results using ns-3 end-to-end simulations.
\section{Simulation Results and Discussion}
\label{sec:SIMULATION RESULTS AND DISCUSSION}
\subsection{Mean throughput}
We use the mean downlink throughput of both technologies as a performance metric for our analyses of the \mbox{Wi-Fi} 6E and 5G \mbox{NR-U} coexistence.
 Each simulation duration is 4 s long, as given in Table \ref{tab:parameters}. We calculate the mean throughput for each AP/gNB over 100 realizations for each network size.

\begin{figure}[htbp]
	\centering
	\includegraphics[width=0.4\textwidth,trim = 90mm 61mm 95mm 40mm,clip]{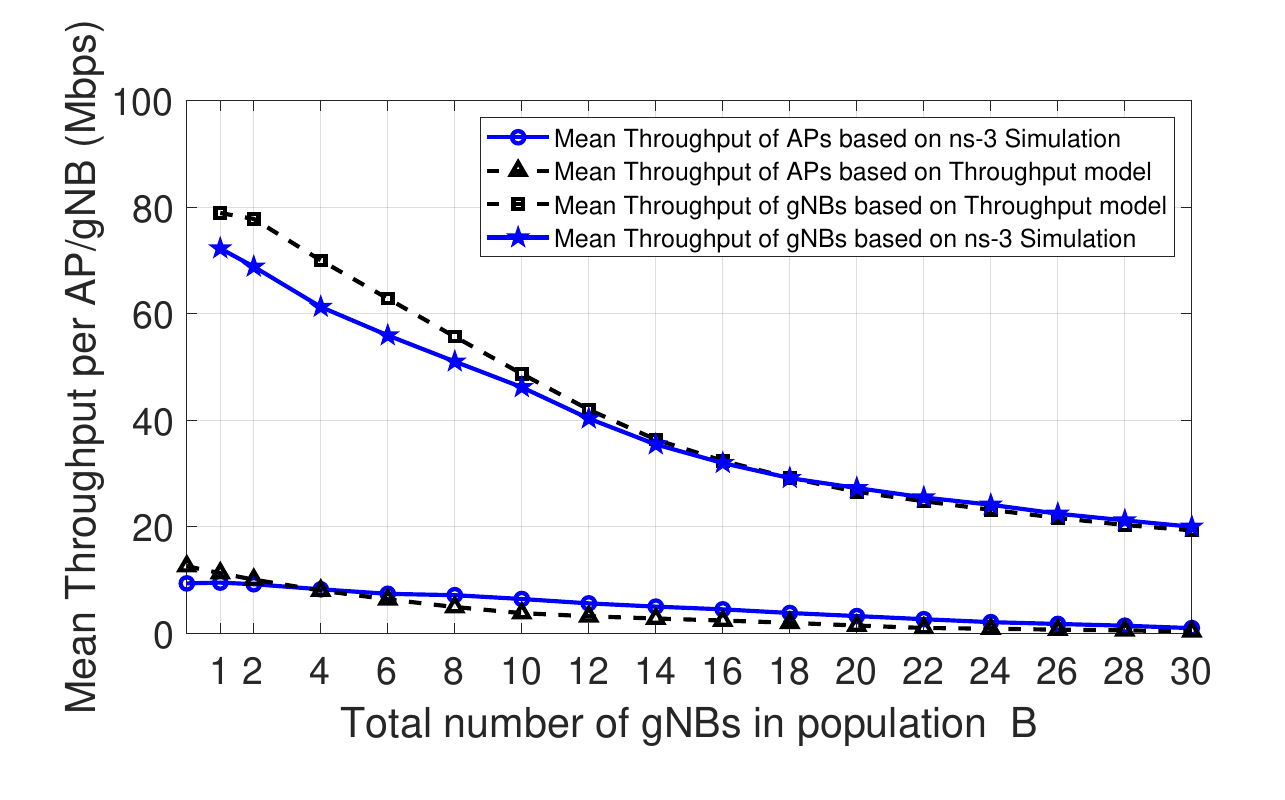}
	\caption{Mean throughput per AP/gNB, while {No. of APs: 10 \& No. of gNBs: [0-30]} and $\Delta$ is set to 1000$\mu s$}
	\label{fig:analyt_sim}
\end{figure}

In Figure \ref{fig:analyt_sim}, we show the mean throughput per AP/gNB obtained from the throughput model and the ns-3 simulation. 
We observe that the simulated mean throughput of \mbox{Wi-Fi} APs and \mbox{NR-U} gNBs in ns-3 and the one obtained by the throughput model elaborated in Section \ref{Throughput_Model_sec} closely match. 

The slight difference in case of gNB throughput can be attributed to two facts. First, the distinct rate control algorithms, and second, the time granularity employed in each approach. In the case of rate control, ns-3 uses a dynamic rate control algorithm that adjusts rates in real-time based on feedback from the environment. However, in the throughput model, the rate is decided based on the SINR-rate mapping, as outlined in the standards. This leads to more optimistic throughput.
In the case of the time granularity, while ns-3 captures per-packet time granularity and changes the rate accordingly based on the received SINR, the throughput model utilizes time-averaged per-link parameters, including airtime ($AirTime_x$) and MAC efficiency ($S_x$), and adapts the rate based on the average calculated SINR. This causes slight differences in throughput from ns-3 and the model, particularly in sparse scenarios due to statistical considerations.



\begin{figure*}[!ht]
	\centering
	\subfigure{
	\includegraphics[width=0.315\textwidth,trim = 111mm 65mm 110mm 50mm,clip]{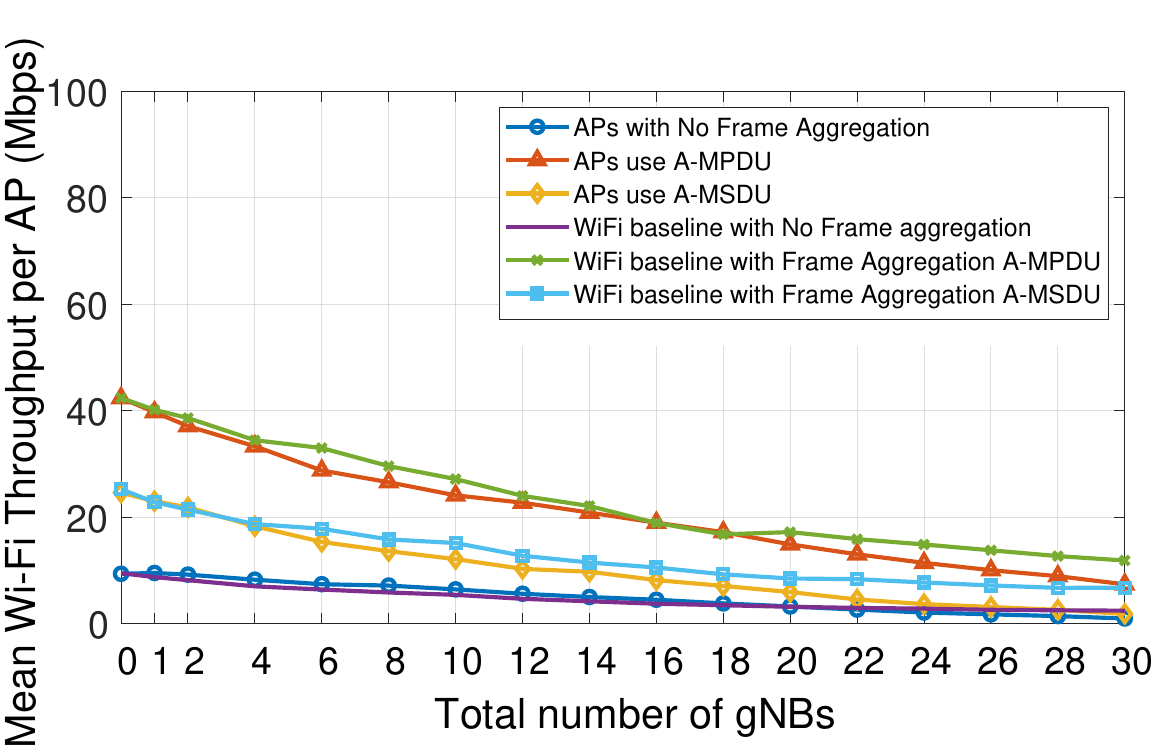}
		\label{TSUS_1_1}	
	}
    \addtocounter{subfigure}{-3}
	\subfigure{
	\includegraphics[width=0.315\textwidth,trim = 115mm 66mm 103mm 50mm,clip]{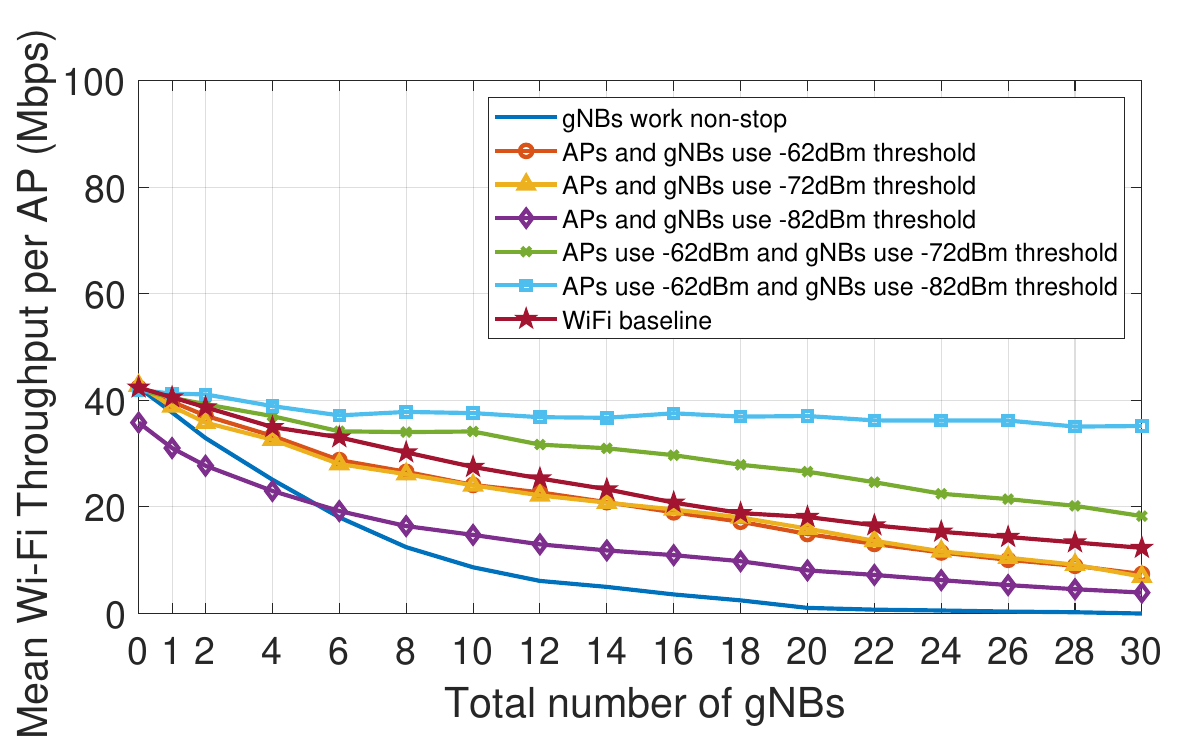}
		\label{TSUS_1_2}	
	}
	\subfigure{
	\includegraphics[width=0.310\textwidth,trim = 0mm 69mm 0mm 57mm,clip]{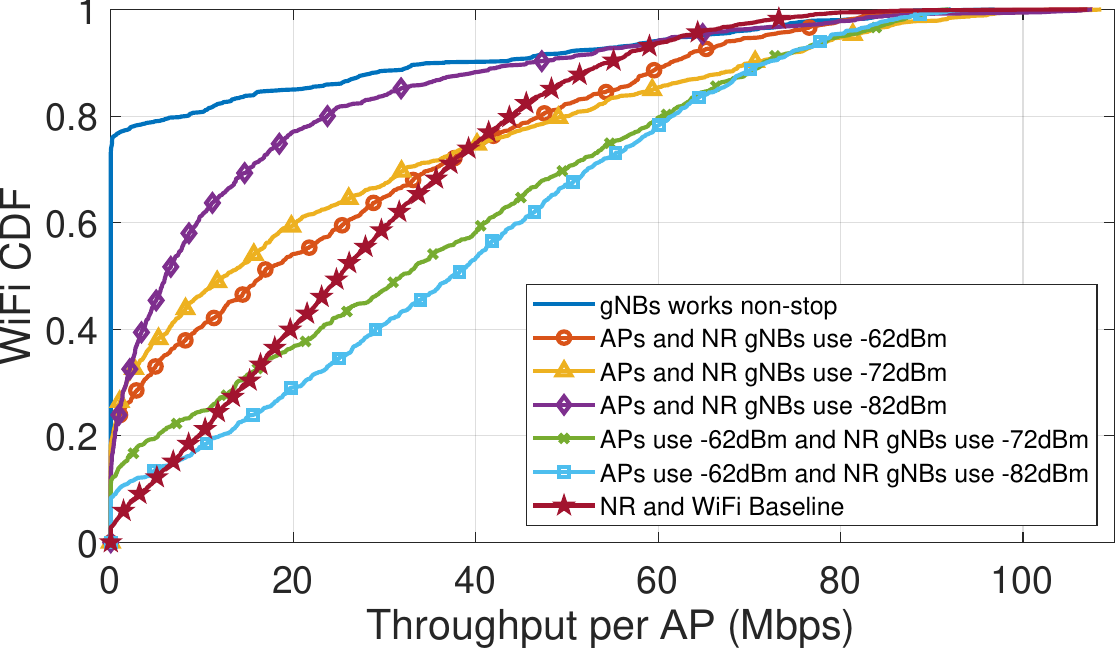}
		\label{TSUS_1_3}	
	}
	\label{nanay}
 \vspace{-0.6cm}
\end{figure*}
\begin{figure*}[!ht]
	\centering
	\subfigure[{No. of APs: 10 \& No. of gNBs: [0-30]}]{
		\includegraphics[width=0.315\textwidth,trim = 113mm 63mm 110mm 55mm,clip]{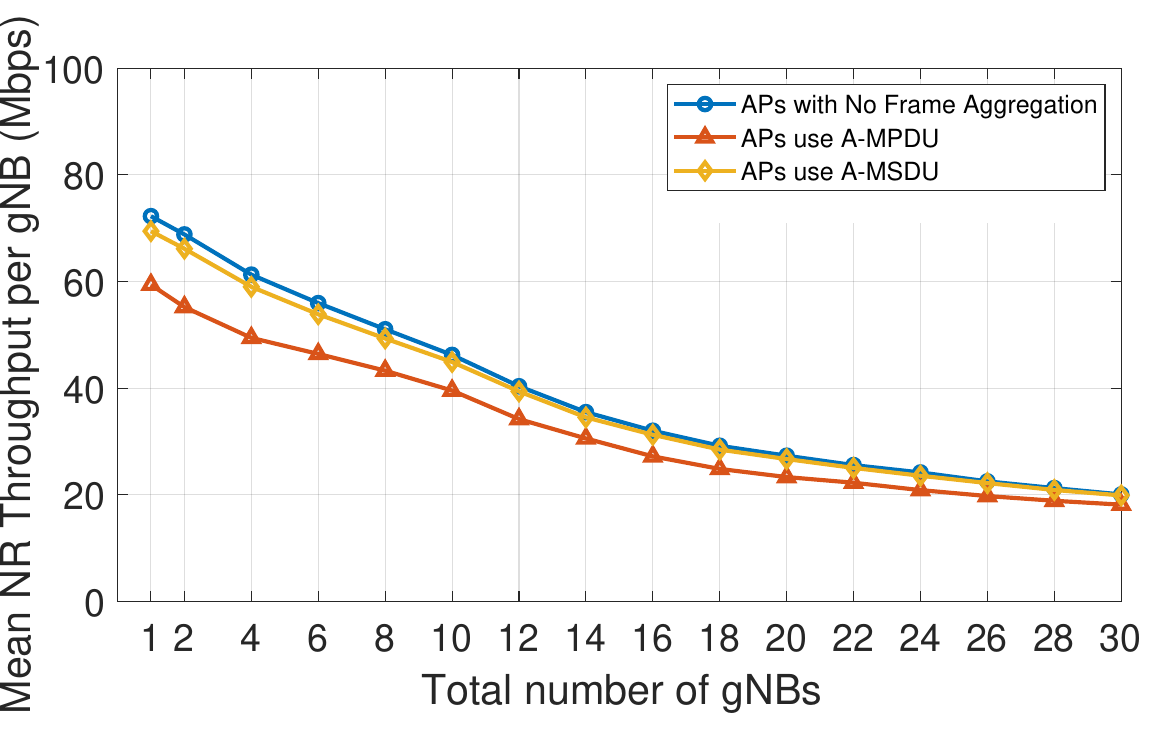}
		\label{TSUS_3_1}	
	}
	\subfigure[{No. of APs: 10 \& No. of gNBs: [0-30]}]{
	\includegraphics[width=0.317\textwidth,trim = 120mm 60mm 98mm 60mm,clip]{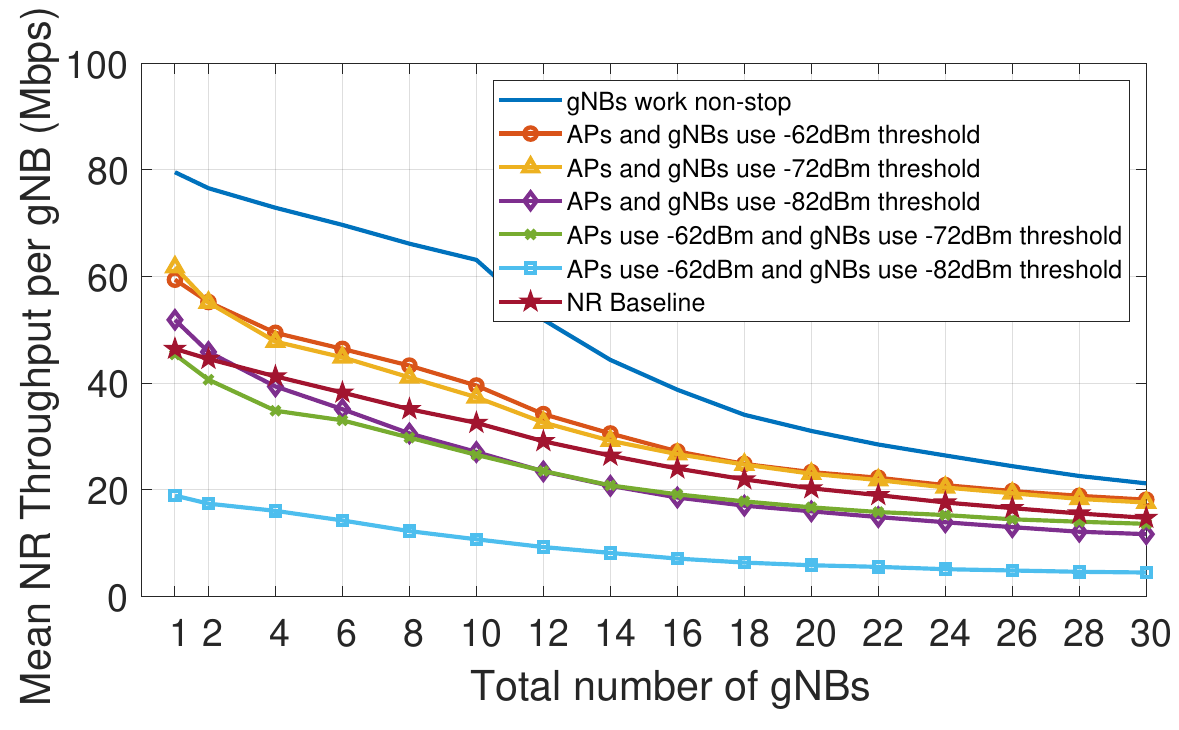}
		\label{TSUS_3_2}	
	}
	\subfigure[No. of APs: 10 \& No. of gNBs: 10]{
	\includegraphics[width=0.310\textwidth,trim = 0mm 31mm 0mm 92mm,clip]{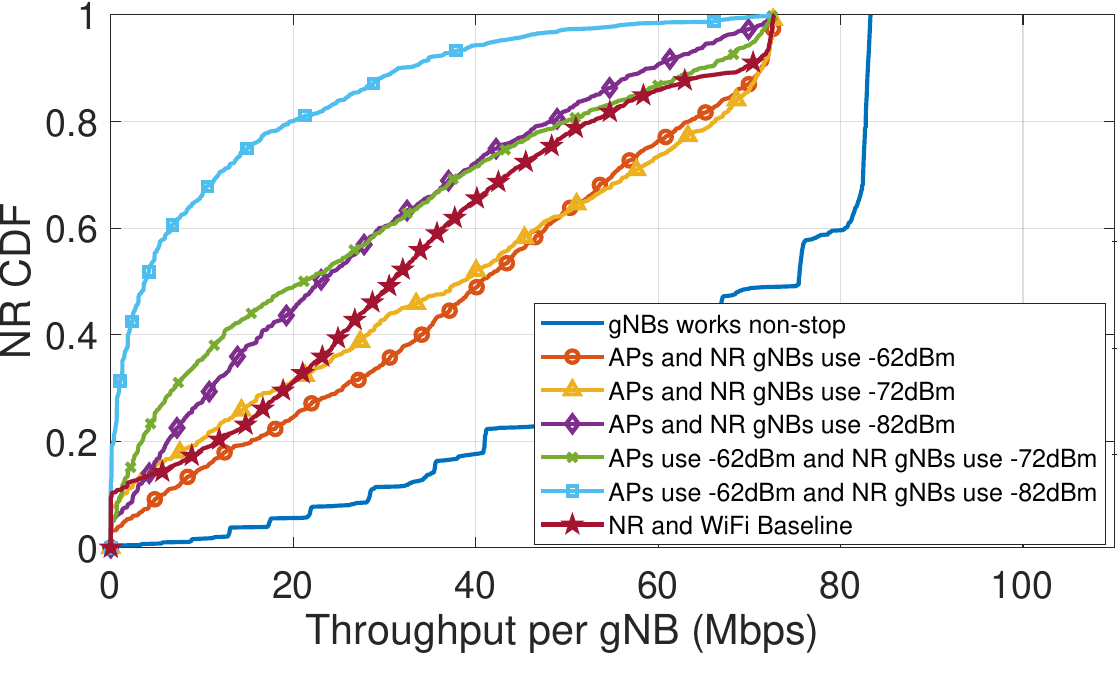}
		\label{TSUS_3_3}	
	}
	\caption{Effect of NR-U and Wi-Fi MAC parameters on mean throughput : (a) Different types of frame aggregation, (b) Different ED threshold, (c) \mbox{NR-U} and Wi-Fi throughput distribution at different thresholds}
	\label{nanay}
\end{figure*}
\subsection{The impact of different types of frame aggregation on the throughput}
In this subsection, we report on how different frame aggregating methods affect both technologies' mean throughput. Figures ~\ref{TSUS_1_1} show the mean throughput per AP and gNB. 
As discussed in section~\ref{sec:ScenariosandSetup}, we consider the maximum available size of A-MPDU and A-MSDU specified in the standard \cite{Committee2007} in our simulations. The \mbox{Wi-Fi} baseline in Figure~\ref{TSUS_1_1} refers to a scenario in which only \mbox{Wi-Fi} devices are deployed (adding \mbox{Wi-Fi} APs rather than 0-30 \mbox{NR-U} gNBs). When frame aggregation is turned off, each AP detects the channel before transmitting each frame. If a gNB identifies an idle channel, it may occupy the channel for the period of MCOT. For example, with 10 gNBs and 10 APs, the \mbox{NR-U} achieves a mean throughput of 46 Mbps per gNB, whereas \mbox{Wi-Fi} achieves a throughput of 6.4 Mbps per AP, as illustrated in Figure~\ref{TSUS_3_1}. When only A-MSDU is used, in which several payload units and one MAC header are packed into a single frame, each AP occupies the channel for a longer length. This reduces the number of \mbox{Wi-Fi} device sensing attempts, improves channel utilization, and enhances \mbox{Wi-Fi} throughput while having a negligible impact on \mbox{NR-U} throughput. In the case of ten gNBs and ten APs, the average throughput of \mbox{Wi-Fi} devices nearly doubles compared to scenarios without frame aggregation.  
 When A-MPDU is enabled for \mbox{Wi-Fi}, the maximum transmission duration (5.484 ms) can be reached, several times longer than when only A-MSDU is used. 
 For example, for 10 gNBs and 10 APs, the mean throughput of \mbox{Wi-Fi} per AP improves from 6.4 Mbps to 24.1 Mbps, but for \mbox{NR-U}, it decreases from 46.2 Mbps to 39.5 Mbps. 
 
By analyzing the results from the frame aggregation technique to the baselines, we observe that using A-MPDU has a significant impact on the fairness between \mbox{Wi-Fi} and \mbox{NR-U} and can improve the \mbox{Wi-Fi} performance. 

\subsection{The influence of different ED thresholds on the throughput}
Figure~\ref{TSUS_3_2} shows \mbox{Wi-Fi} and \mbox{NR-U} mean throughput for different sensing thresholds. We also include two baselines for comparison: one with only \mbox{Wi-Fi} APs (rather than 0-30 gNBs) and the other with only \mbox{NR-U} gNBs (instead of 10 deployed AP). As neither \mbox{Wi-Fi} nor \mbox{NR-U} is the incumbent user in 6 GHz bands, these baselines are plotted. So, comparing the results with one-technology scenarios may help establish if two technologies can coexist fairly. We also present a hypothetical scenario with always-ON gNBs that do not use LBT (Dark blue curve). It should be noted that \mbox{Wi-Fi} APs use A-MPDU for frame aggregation in all cases.
From Figure~\ref{TSUS_3_2}, we observe the following. When gNBs and \mbox{Wi-Fi} APs use -62 dBm as the ED threshold, the gNBs get higher mean throughput. This is because each gNB occupies the channel for 8 ms (MCOT) while \mbox{Wi-Fi} AP keeps the channel for a maximum of 5.484 ms. In addition, \mbox{Wi-Fi} APs defer to other APs with a sensing threshold of a minimum \mbox{-82 dBm}, which makes the APs to be more sensitive compared to the gNBs.
If the sensing threshold for gNBs and APs are set to -72 dBm, the results are almost similar to the case with a sensing threshold of -62 dBm -- only the gNBs experience a slight decrease in throughput due to higher sensitivity compared to -62 dBm threshold scenario.

The results demonstrate that gNB and \mbox{Wi-Fi} mean throughput reduces when the sensing threshold is -82 dBm. Due to their improved sensitivity, they can detect AP and gNB signals from even further away. Thus, they must share the channel with additional users. In all three scenarios with common ED thresholds for APs and gNBs, \mbox{NR-U} gets higher throughput as demonstrated in Figure~\ref{TSUS_3_2}, even with a detecting threshold of -82 dBm, which is the minimum threshold that APs use to defer to other APs.
This is because the maximum transmission duration of the two technologies is not the same. Taking into account the scenario with 10 APs and 10 gNBs, the average throughput of \mbox{NR-U} and \mbox{Wi-Fi} at -62 dBm and -72 dBm is almost the same with 24 Mbps and 38 Mbps, respectively. The mean throughput rates for ED value of \mbox{-82 dBm} are 14.7 Mbps and 26.5 Mbps, respectively, substantially lower than the simulation with -62 dBm and -72 dBm thresholds. This illustrates that the -82 dBm threshold wastes channel bandwidth without affecting throughput fairness.

Finally, we consider scenarios when the ED thresholds for \mbox{Wi-Fi} APs and \mbox{NR-U} gNBs are asymmetric. The \mbox{NR-U} threshold is adjusted to -72 and -82 dBm, whereas the \mbox{Wi-Fi} threshold remains at -62 dBm. We observe that \mbox{Wi-Fi} throughput remains virtually constant at roughly 40 Mbps for gNB's ED threshold value of -82 dBm, since \mbox{NR-U} gNBs are now more sensitive and compete with more devices, resulting in decreased throughput. Here, \mbox{Wi-Fi} signals dominate the building. If the sensing threshold of \mbox{NR-U} is adjusted to \mbox{-72 dBm}, \mbox{Wi-Fi} throughput remains slightly superior to \mbox{NR-U} throughput. But, compared to results with same thresholds for both technologies, \mbox{NR-U} and \mbox{Wi-Fi} throughput are closest. 
For example, Figure~\ref{TSUS_3_3} shows the distribution of throughput per AP/gNB when 10 APs and 10 gNBs are placed in the building. Compared to the NR and \mbox{Wi-Fi} baseline, \mbox{Wi-Fi} has higher mean throughput if the ED threshold of gNBs is just modified. The scenario with -62 dBm and -72 dBm ED thresholds for both APs and gNBs and the scenario with -62 for APs and -72 for gNBs are closer to the baselines. To that end, we can conclude that the preceding scenarios for the ED threshold are appropriate for achieving fairness.

\subsection{The impact of MCOT on coexistence}
\label{MCOT_impact}
\begin{figure}[htbp]
	\centering
	\includegraphics[width=0.4\textwidth,trim =  105mm 70mm 110mm 45mm,clip]{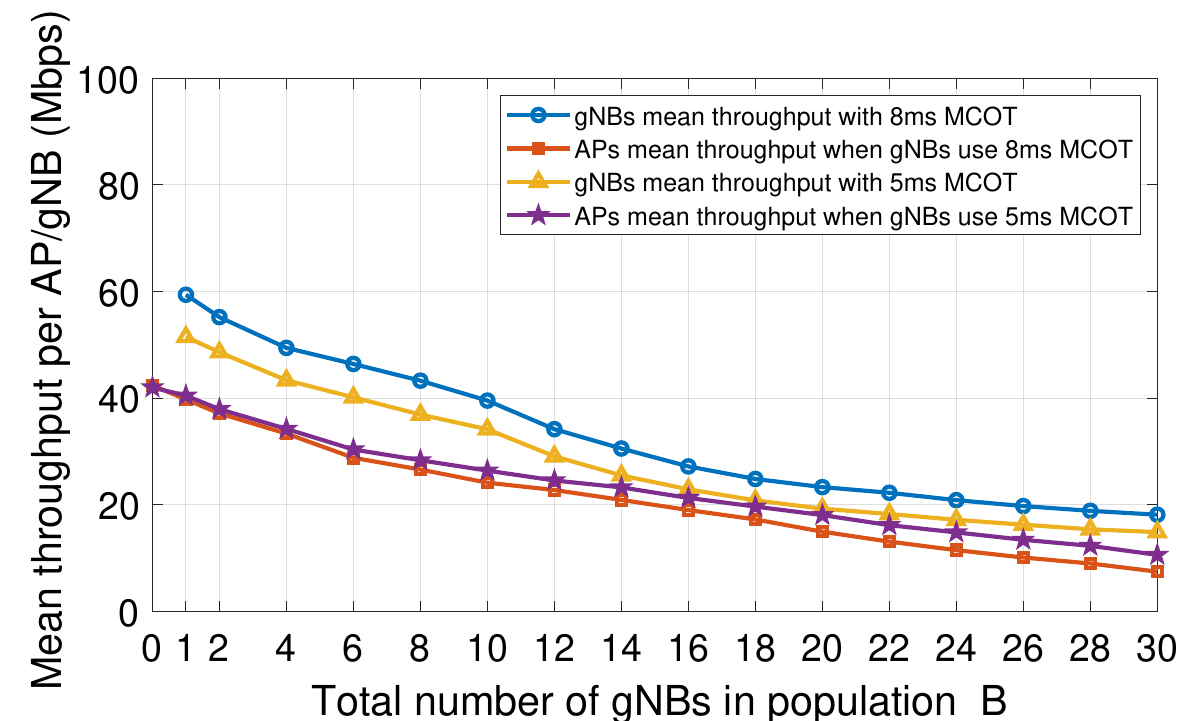}
	\caption{{No. of APs: 10 \& No. of gNBs: [0-30]}, with frame aggregation A-MPDU, \mbox{5} and \mbox{8 ms} MCOT}
	\label{fig:10}
\end{figure}

In this subsection, we study the duration of MCOT on the coexistence. We run the simulations for MCOT of 5 ms and ED threshold of -62 dBm for both \mbox{Wi-Fi} and \mbox{NR-U}. Figure~\ref{fig:10} compares the mean throughput per AP/gNB for MCOT values of 8 ms and 5 ms. Note that we use an MCOT value of \mbox{5 ms} for our simulation because it is near to the maximum transmission duration of a \mbox{Wi-Fi} 6E AP. 
When MCOT is set to 5 ms, the results show only a small increment in \mbox{Wi-Fi} throughput. The mean throughput of \mbox{NR-U} decreases, which is to be expected given the decreased MCOT. For example, with 10 gNBs and 10 APs, the mean throughput per AP increases by only 2.2 Mbps, while the mean throughput per gNB decreases by 5.4 Mbps. The reason for this is even with 5 ms MCOT, each AP has to compete with other AP/gNBs to get the channel, and \mbox{Wi-Fi} APs are more sensitive because \mbox{Wi-Fi} AP uses -82 dBm to defer to other \mbox{Wi-Fi} APs, while gNBs still use -62 dBm detection threshold for both technologies.
\begin{figure*}[!ht]
	\centering
	\subfigure[{No. of APs: 10 \& No. of gNBs: [0-30]}]{
		\includegraphics[width=0.315\textwidth]{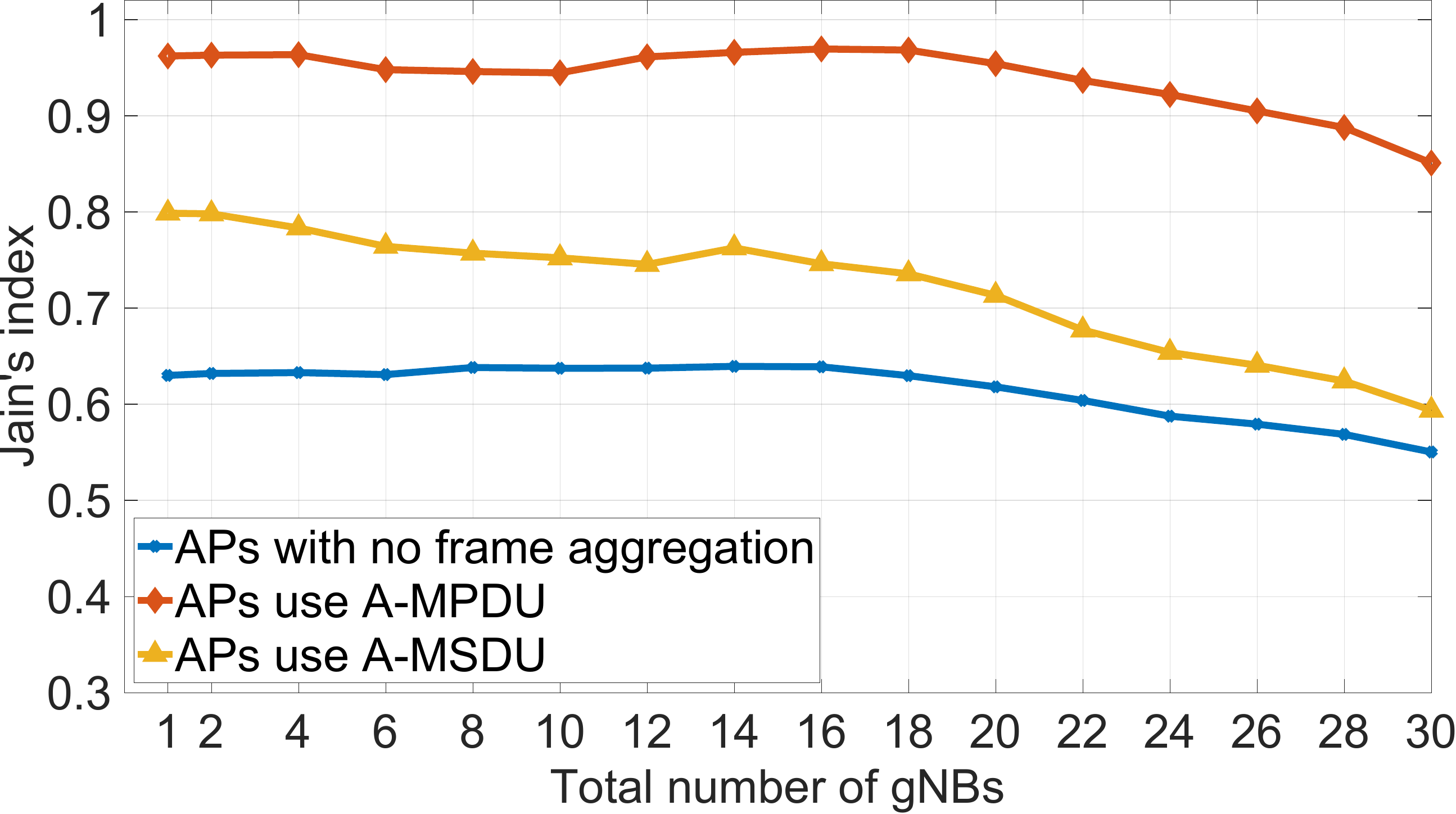}
		\label{TSUS_2_1}	
	}
	\subfigure[{No. of APs: 10 \& No. of gNBs: [0-30]}]{
		\includegraphics[width=0.315\textwidth]{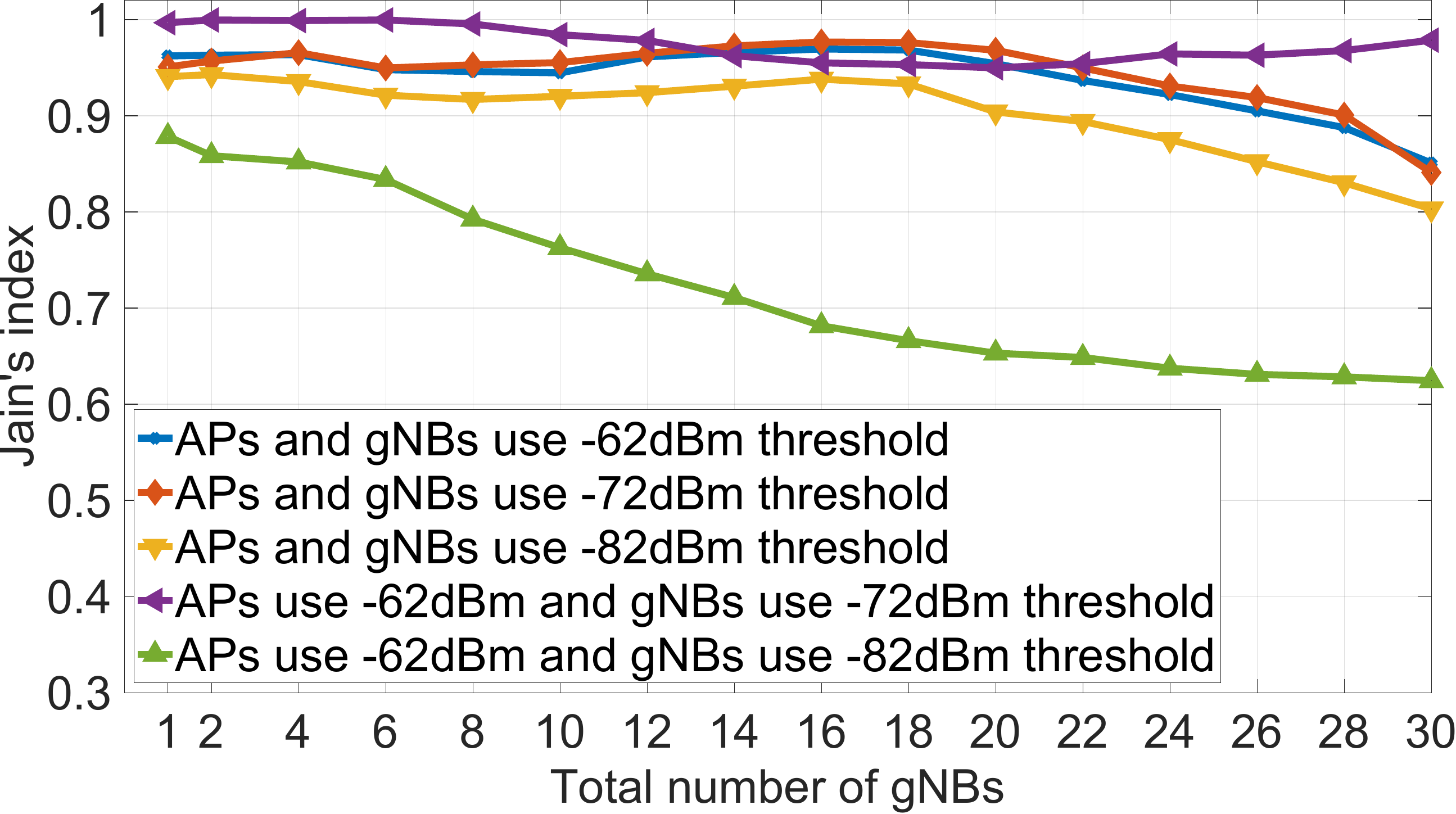}
		\label{TSUS_2_2}	
	}
	\subfigure[{No. of APs: 10 \& No. of gNBs: [0-30]}]{
		\includegraphics[width=0.315\textwidth]{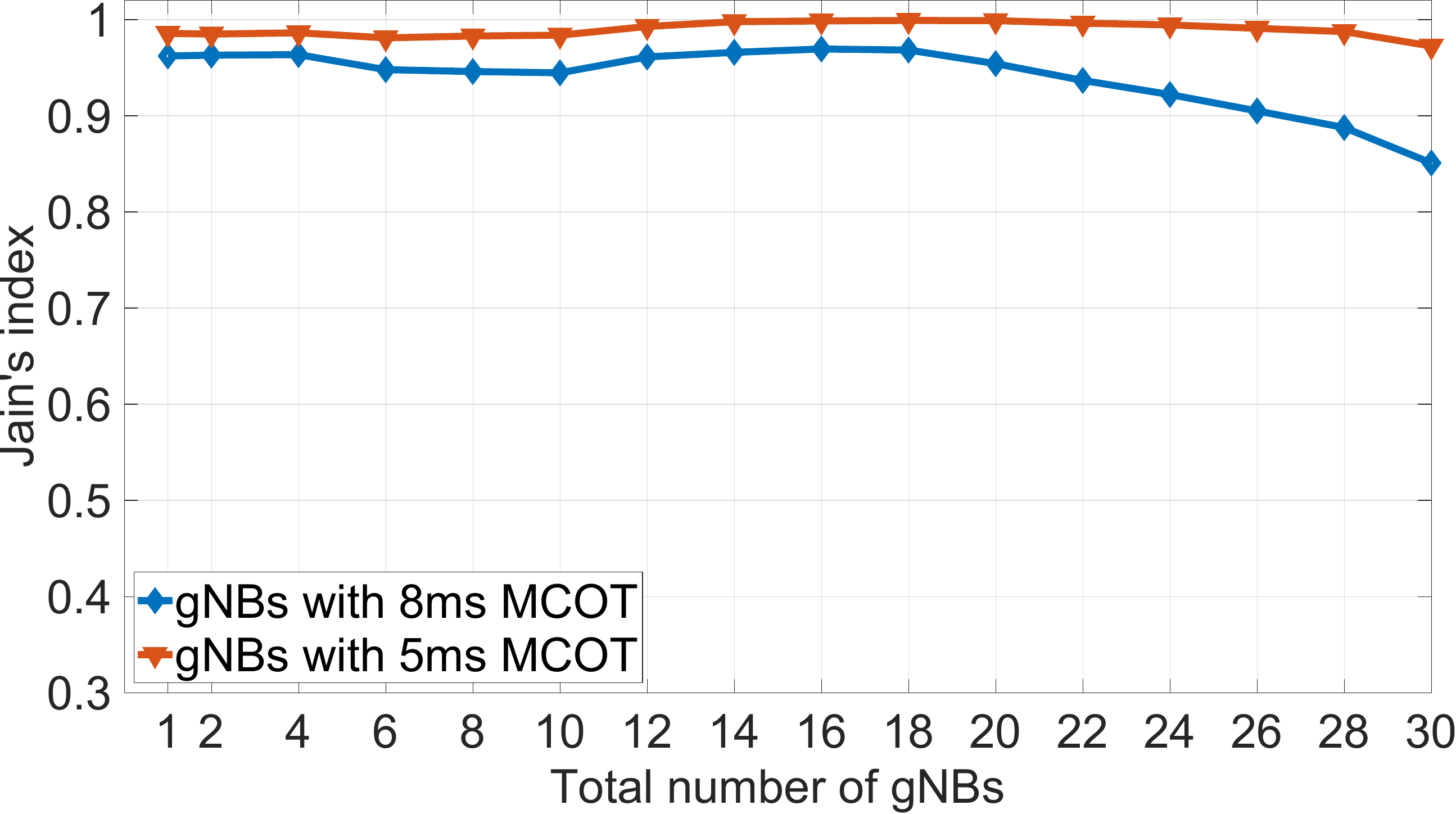}
		\label{TSUS_2_3}	
	}
	\caption{Effect of NR-U and Wi-Fi MAC parameters on fairness: (a) Different types of frame aggregation, (b) Different ED thresholds, (c) Maximum channel occupancy time}
	\label{jain_index_2}
\end{figure*}
\subsection{Throughput fairness}
\label{result_fairness}
We use Jain's fairness index\cite{Jain} to evaluate the throughput fairness between the networks. The Jain index measures quantitative fairness. The value of Jain's fairness index is between 0 and 1. Fairer network throughput enhances channel access fairness as the index approaches one. Jain's throughput fairness index is calculated using the following equation:
\begin{equation} \label{eqn1}
	J=\frac{[\,\sum_{i=1}^n{x_i}]\,^2}{n\sum_{i=1}^n {x^2_i}}
\end{equation}
where $x_i$ is the mean throughput of $i^{th}$ network and $n$ is the number of networks. 
We observe in Figure \ref {TSUS_2_1} that maximum fairness can be reached when the \mbox{Wi-Fi} uses A-MPDU-based frame aggregation. On the other hand, when \mbox{Wi-Fi} APs do not opt for frame aggregation, the value of the fairness index decreases due to shorter \mbox{Wi-Fi} transmission time in comparison to \mbox{NR-U} MCOT duration.       
In Figure \ref {TSUS_2_2}, we study the effect of the ED threshold on fairness. We see that by using the same -62 dBm and -72 dBm threshold for both \mbox{Wi-Fi} and \mbox{NR-U}, we obtain fairness index value of 0.9. However, the best fair result comes when \mbox{Wi-Fi} uses -62 dBm, and gNBs use -72 dBm. In this case, we get more than a 0.95 fairness index for any number of gNBs. 
Figure \ref {TSUS_2_3} compares scenarios with MCOT values of 8 and 5 ms. The result in Figure \ref {TSUS_2_3} shows that with MCOT of 5 ms, Jain’s fairness index value close to 1 can be reached. 

By analyzing the fairness curves in all three subfigures, we observe that the inter-technology fairness can be satisfied by changing the gNBs detection threshold to -72 dBm, while \mbox{Wi-Fi} APs enable A-MPDU frame aggregation. 
\section{Conclusions}
\label{Conclusions}
In this paper, we performed analytical and simulation-based studies on the coexistence of \mbox{Wi-Fi} 6E and 5G \mbox{NR-U} in terms of mean achievable downlink throughput and fairness. We furthermore explored how different parameter settings of several key features, such as the MAC frame aggregation, the energy detection thresholds, and MCOT affect the throughput performance and fairness when both technologies are contending for the same channel. Our results show that enabling frame aggregation in the \mbox{Wi-Fi} APs ensures better channel utilization and higher throughput when coexisting with \mbox{NR-U}. The study on the different ED threshold levels revealed that using a common threshold of -62 dBm and -72 dBm or -62 dBm for AP and -72 dBm for gNBs provides the most harmonious coexistence. We also showed that changing MCOT from 8 ms to 5 ms only slightly increases the \mbox{Wi-Fi} throughput. However, it increases the fairness of the system at the cost of \mbox{NR-U} throughput. Our results underline that selecting the correct ED threshold is more critical for throughput fairness than MCOT in a dense scenario. 
\begin{acks}
The authors acknowledge the financial support by the Federal Ministry of Education and Research of Germany in the project “Open6GHub” (grant number:16KISK012). Simulations were performed with computing resources granted by RWTH Aachen University under project rwth0767.
\end{acks}

\bibliographystyle{ACM-Reference-Format}
\bibliography{sample-base}

\end{document}